\documentclass[]{aastex631}
\usepackage{bm}

\shorttitle{Whistler Wave Growth: MMS Observation}
\shortauthors{He et al.}
\graphicspath{{./}{figures/}}
\begin{document}

\title{Observations of rapidly growing whistler waves in front of space plasma shock}

\author[0000-0001-8179-417X]{Jiansen He}
\affiliation{School of Earth and Space Sciences, Peking University, Beijing 100871, Beijing, China}

\author[0000-0002-1541-6397]{Xingyu Zhu}
\affiliation{School of Earth and Space Sciences, Peking University, Beijing 100871, Beijing, China}

\author[0000-0001-7697-1186]{Qiaowen Luo}
\affiliation{School of Atmospheric Sciences, Sun Yat-sen University, Zhuhai, 519000, China}

\author[0000-0001-7205-2449]{Chuanpeng Hou}
\affiliation{School of Earth and Space Sciences, Peking University, Beijing 100871, Beijing, China}

\author[0000-0002-0497-1096]{Daniel Verscharen}
\affiliation{Mullard Space Science Laboratory, University College London, Dorking RH5 6NT, UK}

\author[0000-0002-6300-6800]{Die Duan}
\affiliation{School of Earth and Space Sciences, Peking University, Beijing 100871, Beijing, China}

\author[0000-0003-1920-2406]{Wenya Li}
\affiliation{State Key Laboratory of Space Weather, National Space Science Center, Chinese Academy of Sciences, Beijing 100190, China}

\author[0000-0002-3859-6394]{Jinsong Zhao}
\affiliation{Purple Mountain Observatory, Chinese Academy of Sciences, Nanjing 210008, China}

\author[0000-0003-3072-6139]{Tieyan Wang}
\affiliation{RAL Space, Rutherford Appleton Laboratory, Harwell Oxford, Didcot OX11 0QX, UK}

\author[0000-0002-1046-746X]{Daniel B. Graham}
\affiliation{Swedish Institute of Space Physics, Uppsala SE-75121, Sweden}

\author[0000-0002-6414-3794]{Qiugang Zong}
\affiliation{School of Earth and Space Sciences, Peking University, Beijing 100871, Beijing, China}

\author[0000-0001-6826-2486]{Zhonghua Yao}
\affiliation{Key Laboratory of Earth and Planetary Physics, Institute of Geology and Geophysics, Chinese Academy of Sciences, Beijing 100029, China}

\begin{abstract}
Whistler mode wave is a fundamental perturbation of electromagnetic fields and plasmas in various environments including planetary space, laboratory and astrophysics. The origin and evolution of the waves are a long-standing question due to the limited instrumental capability in resolving highly variable plasma and electromagnetic fields. Here, we analyse data with the high time resolution from the multi-scale magnetospheric spacecraft in the weak magnetic environment (i.e., foreshock) enabling a relatively long gyro-period of whistler mode wave. Moreover, we develop a novel approach to separate the three-dimensional fluctuating electron velocity distributions from their background, and have successfully captured the coherent resonance between electrons and electromagnetic fields at high frequency, providing the resultant growth rate of unstable whistler waves. Regarding the energy origin for the waves, the ion distributions are found to also play crucial roles in determining the eigenmode disturbances of fields and electrons. The quantification of wave growth rate can significantly advance the understandings of the wave evolution and the energy conversion with particles. 
\end{abstract}

\section{Introduction} \label{sec:intro}
The emission, propagation, and dissipation of plasma waves are fundamental processes in the planetary space, the interplanetary space, and beyond \citep{Le2013,Narita2018,Shan2020,Zong2020,Burch2016,Howes2017,Bruno2013,Verscharen2019}, and are ranked as one of key objectives by various space exploration programs \citep{Burch2016,Fox2016}. The growth and dissipation of wavelike turbulences are key consequences of energy conversion between fields and particles, which have been a research focus of the space physics community for decades \citep{Stix1962}. The growth of waves could lead to the formation of shocklet or shock structures \citep{Tsurutani1989}. In previous literature, the instability for plasma waves is estimated by calculating the growth rate according to linear plasma theory with a pre-defined background plasma state and magnetic field as the observational input \citep{Bale2009,Zhao2019}. A direct examination of wave growth rate from spacecraft observations is crucial to validate the theoretical studies. The key barrier to obtain direct observational evidence of wave growth is the limited resolving capability of space instruments, which have a breakthrough due to the successful launch of MMS (Magnetospheric Multiscale) mission. MMS provides state-of-the-art high quality measurements of electromagnetic fields and of the particle species by the FIELDS \citep{Torbert2016} and FPI \citep{Pollock2016} instruments, respectively.

The endeavors dedicated to diagnosing field-particle interactions have been recently arousing wide attention and significant concerns. Landau damping of kinetic Alfv\'en waves is highlighted by the observed correlation between the electron velocity distribution function and the parallel electric field \citep{Gershman2017,Chen2019}. Phase relations between wave fields and ion differential energy flux were observed, and the energy is suggested to transfer from ions to ion cyclotron waves \citep{Kitamura2018}. The spectra of turbulence dissipation rate were provided for ion cyclotron waves and kinetic Alfv\'en waves in magnetosheath turbulence \citep{He2020}. However, these recent investigations are limited to low frequencies, and have never exceeded the ion gyro-frequency. The growth process whistler waves at higher frequencies, which are ubiquitous and crucial for understanding high-frequency kinetic physics in multiple plasma environments \citep{Wilson2013,Stansby2016}. For whistler waves, the correlated change of the electron velocity distribution has not been presented before. Its phase coherence with the perturbations of electromagnetic field is also yet to be confirmed.      
We target the foreshock region to investigate the growing process of whistler waves, since the plasma number density is sufficiently high for a detection at the time scale of electron gyro-motion and the magnetic field is weak enough for the gyro-motion to be captured by the detector recording. As a result, we reveal the comprehensive process of the field-particle interaction at a high measurement cadence, manifesting the ongoing rapid growth of whistler waves, which finally evolve to large amplitudes and contribute to shocklet formation. The field-particle coupling for whistler waves of this work is a crucial constituent of the collisionless shock physics, especially for the energy deposit and conversion of a supercritical shock.

\section{Observational Analysis}
\subsection{Observation of Whistler Waves as Precursor Signal in Foreshock}

\begin{figure}
    \centering
    \includegraphics[width=12cm]{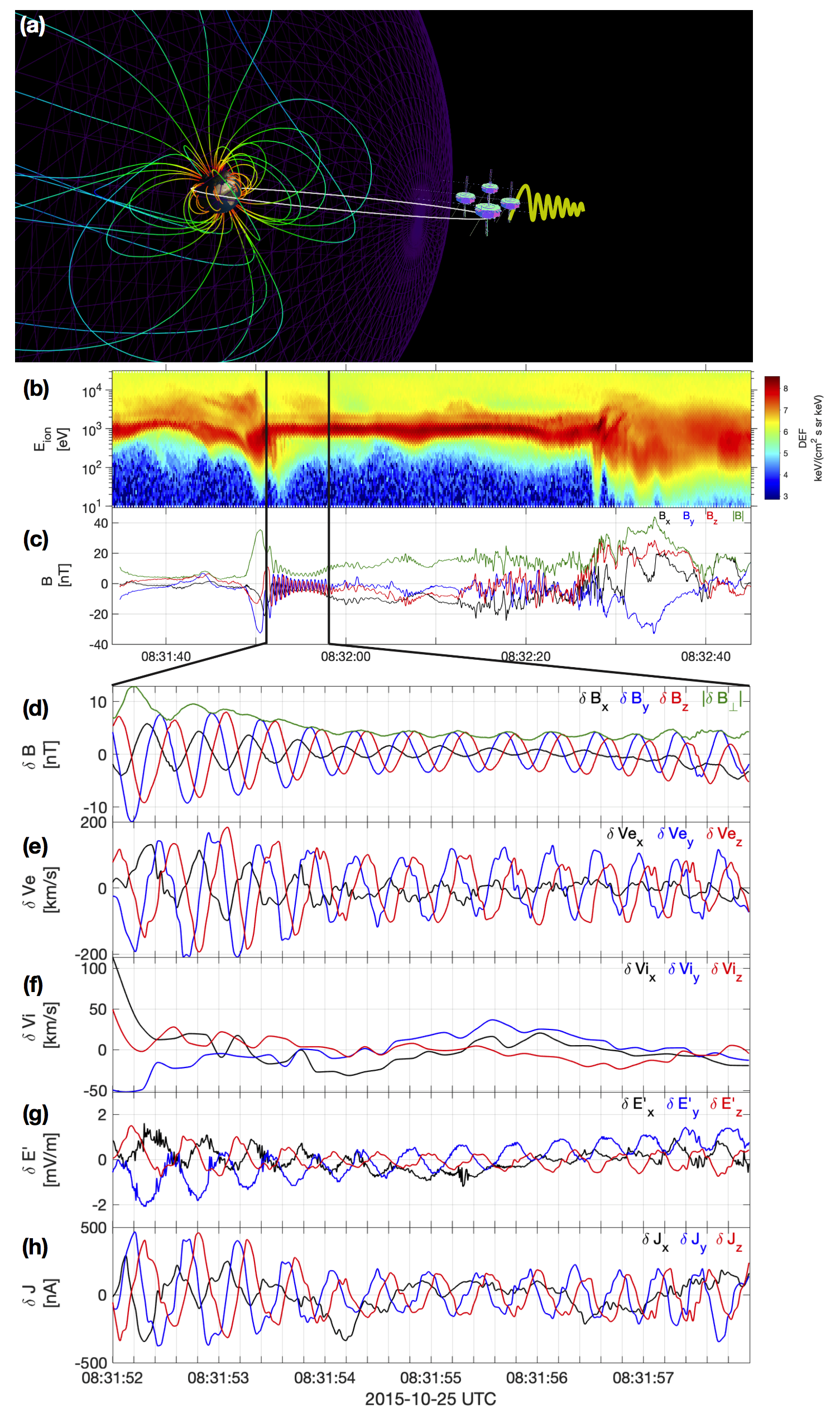}
    \caption{The MMS spacecraft outside Earth's bowshock observe an event of large-amplitude whistler waves. (a) Sketch of particles and waves measured by MMS in front of the bow shock. The two “+” circles represent the solar wind protons and shock-reflected protons. The yellow lines denote interplanetary magnetic field lines according to Parker spiral model. (b) Ion energy spectrum of differential flux density in a longer interval including the event marked by the two vertical lines. (c) Time sequences of magnetic field components in a longer interval. (d) Oscillations of the magnetic field ($\delta \bm{B}$). (e) Correlated oscillations of the electron fluid velocity ($\delta \bm{V_{\rm e}}$). (f) Weak oscillations of the ion fluid velocity ($\delta \bm{V_{\rm i}}$). (g) Oscillations of the electric field in the ion mean bulk flow frame ($\delta \bm{E'}$). (h) Oscillations of the current density ($\delta \bm{J}$). }
    \label{fig1}
\end{figure}

\begin{figure}
    \centering
    \includegraphics[width=18cm]{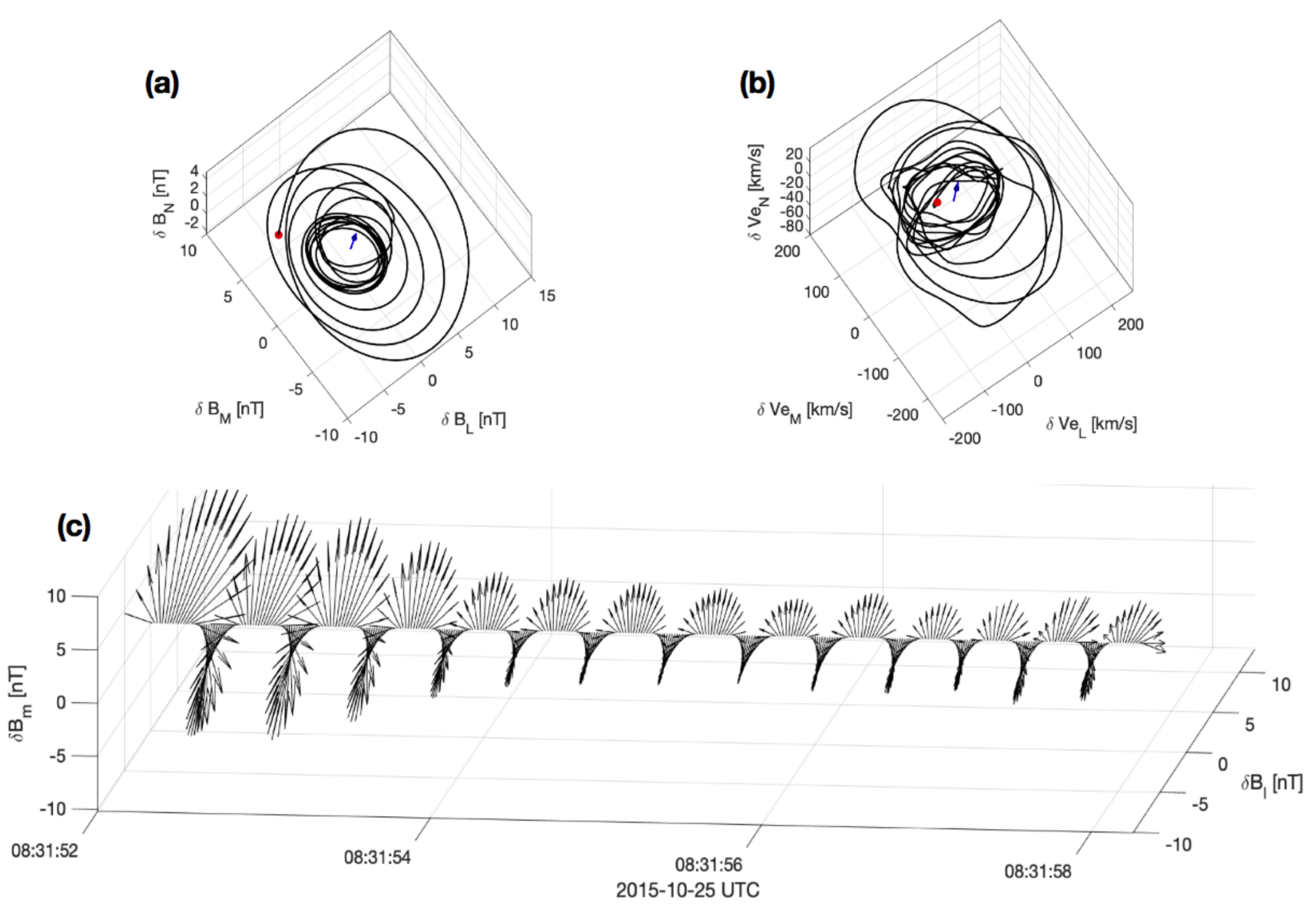}
    \caption{Magnetic field and electron fluid velocity disturbances show right-handed polarization which is typical for whistler waves. (a \& b) Hodograms of the fluctuating magnetic field vectors ($\delta \bm{B}$) and the electron fluid velocity vectors ($\delta \bm{V_{\rm e}}$) in LMN coordinates as transformed from GSE coordinates through MVA. (c) Arrangement of fluctuating magnetic field vector ($\delta \bm{B}$) in chronological order.}
    \label{fig2}
\end{figure}

MMS travelled in the upstream region of the terrestrial bow shock at 08:31 UT on 2015-10-25 (Figure \ref{fig1}a). Supercritical shocks like the Earth’s bow shock are typical in the heliosphere when the supersonic solar wind is hindered by a planetary magnetosphere, forming foreshocks in their upstream regions \citep{Eastwood2005,Turner2020}. Burst-mode measurements of the fields and particles with high quality and high cadence provide an opportunity to reveal the process of field-particle interactions. In Figure \ref{fig1}, we show the wave signatures of different measurement variables. We note that the wave activity is adjacent to a magnetic pulse structure likely to be a shocklet/short large-amplitude magnetic structure (SLAMS) (see Figure \ref{fig1}b \& \ref{fig1}c), which is a typical structure in the foreshock and speculated to nonlinearly evolve from oblique whistler waves \citep{Wilson2016}. The magnetic field vectors oscillate in phase with the electron bulk velocities (Figure \ref{fig1}d \& \ref{fig1}e), and both of them rotate quasi-circularly in the L-M plane in LMN coordinates (Figure \ref{fig2}) (see Methods section for the coordinates definition). On the contrary, the ions’ bulk velocity exhibits a much weaker oscillation leaving the current density oscillation mainly contributed by electrons’ bulk velocity (Figure \ref{fig1}f). The electric field vectors also experience quasi-periodic oscillations (Figure \ref{fig1}g). Note that, we use $-\bm{V}_{\rm e}\times\bm{B}$ to approximate the $E_{\rm z}$ component since the quality of $E_{\rm z}$ as measured by the Axial Double Probe (ADP) instrument is not as good as $E_{\rm x}$ and $E_{\rm y}$ measured by the Spin-plane Double Probe (SDP) instrument. This approximation would somewhat underestimate the growth rate but does not affect the conclusion. We find the wave propagation in the solar wind flow frame to be in the upwind/sunward direction, based on both the multi-spacecraft timing method \citep{Paschmann1998} and the Singular-Value Decomposition method \citep{Santolik2003}. Therefore, we identify that the wave activity is associated with quasi-parallel right-hand polarized whistler waves, which changes to a left-hand polarization in the spacecraft frame due to Doppler-shift effect. The waves’ relative amplitude is large ($\left|\delta B\right|/B_{\rm 0}>1$), indicating a nonlinear state. Quasi-parallel propagating whistler waves with constant magnitudes of the transverse bulk velocity and magnetic field fluctuations are solutions of the full nonlinear multi-fluid equations for all wave amplitudes \citep{Marsch2011}. Therefore, we can use the theoretical prediction from the linear plasma wave theory to investigate the essential physics behind our observations.

\subsection{Observation of Field-Particle Correlation for the Whistler Waves}
\begin{figure}
    \centering
    \includegraphics[width=16cm]{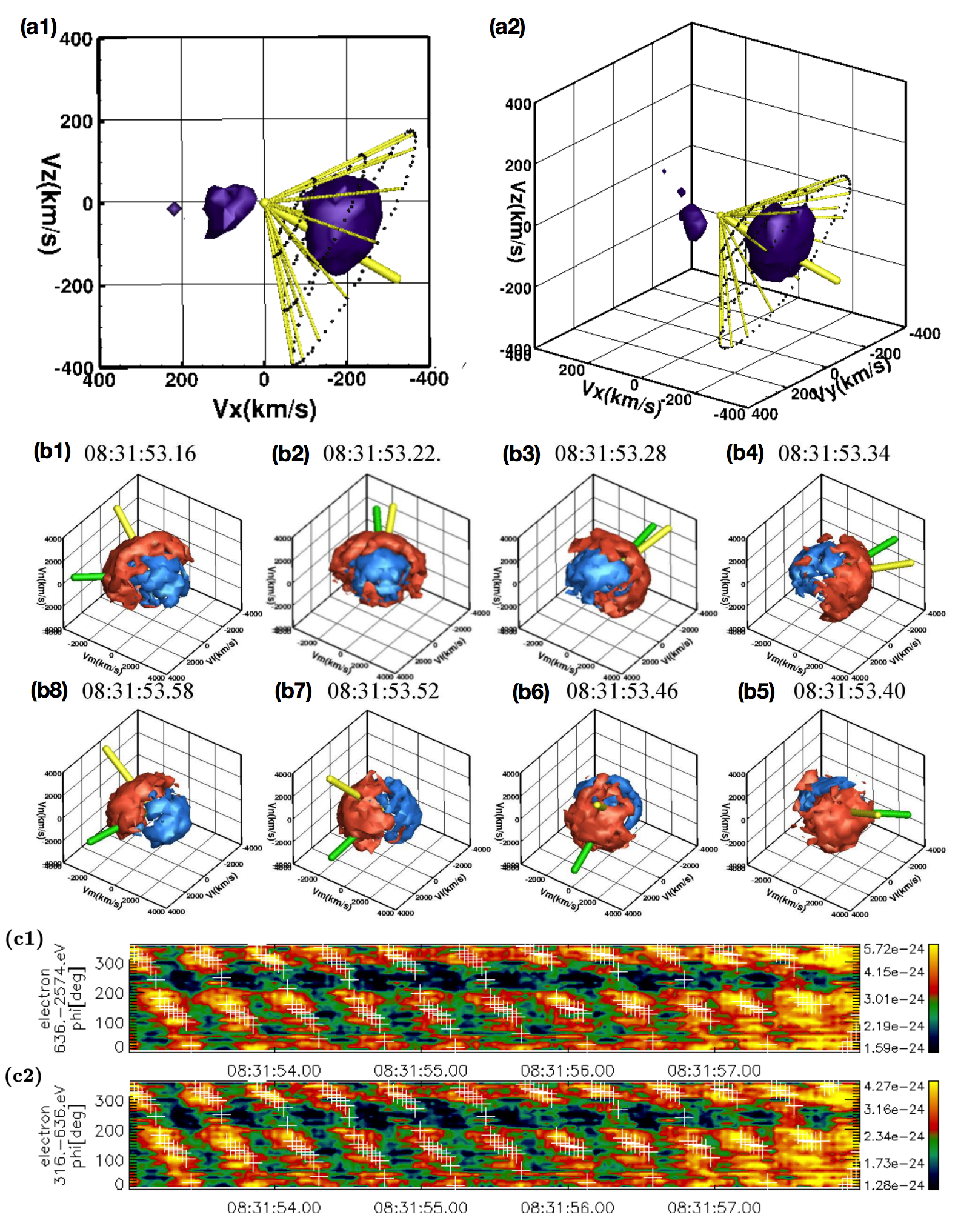}
    \caption{Ion-ion drift velocity distribution as the source of free energy to excite whistler waves. Clear correlations between the electromagnetic wave field and the electron velocity distributions show the field-particle interactions that drive the growth of whistler waves. (a1 \& a2) Ion velocity distribution in the GSE coordinates characterized by a pattern of “core-beam-drift”: the dark purple surfaces on the right and left sides of the origin show the solar wind ions as the core population and the shock-reflected ions as the beam population. The global mean magnetic field vector ($\bm{B}_0$) and the local magnetic field vectors ($\bm{B}_0+\delta \bm{B}$) (disturbed by the wave fluctuations) are denoted by the thick and thin yellow sticks, respectively. (b1-b8) Disturbed electron velocity distributions in the LMN coordinates (red isosurface for $\delta f_{\rm e}>0$, blue isosurface for $\delta f_{\rm e}<0$) rotate in phase with the wave electromagnetic field vectors ($\delta\bm{B}$ by green stick, $\delta \bm{E}'$ by yellow stick) at sub-second periods. (c1-c2) We find a correlation between the azimuthal angle distribution of electron phase space densities (colored background) and the azimuthal angles of the wave electric field vectors (white crosses). }
    \label{fig3}
\end{figure}

In the Earth centered reference frame, the observed ion velocity distribution function ($f_{\rm i}$) consists of two distinct populations: an Earthward core population of solar wind ions and an anti-Earthward beam population of ions reflected from the bowshock (Figure \ref{fig3}a1 \& \ref{fig3}a2). The whistler wave’s fluctuating magnetic field vectors are also illustrated in velocity space, gyrating like the ridges of an umbra around the ion core population. The non-Maxwellian ions are likely to drive plasma instabilities and excite waves. However, before this work, it still remains a long-lasting challenge to capture the growing process of wave activity since the essential field-particle interaction responsible for the energy transfer has yet to be revealed. What would happen to electrons for coupling under the condition of non-Maxwellian ions?

The oscillation of the electron velocity distribution function ($\delta f_{\rm e}$) as a perturbation related to the eigenmode represents the key role of particles in the process of field-particle interaction. We find that positive (red) and negative (blue) $\delta f_{\rm e}$ are located almost opposite to one another and gyrate in velocity space with the same period as that of the fluctuating electromagnetic fields ($\delta \bm{E}'$ (green), $\delta \bm{B}$ (yellow)) (Figure \ref{fig3}b1-\ref{fig3}b8, also supplementary Animation 1). Moreover, the electric field vector always points towards positive $\delta f_{\rm e}$ during the entire wave period, showing that the energy is transferred from the particles to the fields rather than the inverse. The good phase correlation between the $\delta \bm{E'}$'s azimuthal angle ($\phi\left(\delta \bm{E}'\right)$) and the enhanced-$f_{\rm e}$’s azimuthal angle ($\phi\left({\rm enhanced}-f_{\rm e}\right)$) is also illustrated in Figure \ref{fig3}c1 and \ref{fig3}c2.

\subsection{From Field-Particle Correlation to Wave Growth}
The Vlasov equation, which involves the background and disturbed velocity distributions as well as the disturbed electromagnetic fields, can be expressed as
\begin{equation}
    \frac{\partial}{\partial t}(f_0+\delta f)+\frac{q}{m}(\delta\bm{E}+\bm{V}\times\delta\bm{B})\cdot\nabla_V(f_0+\delta f)+\bm{V}\cdot\nabla_X(f_0+\delta f)=0
\end{equation}
where $f_0$, $\delta f$, $\delta\bm{E}$, and $\delta\bm{B}$ are the background velocity distribution, the wave-disturbed velocity distribution, and the wave electric and magnetic fields, respectively. The electric field part in the Lorentz force is responsible for the energy transfer between fields and particles, while the magnetic field part contributes to the energy transfer of the particles themselves between the parallel and perpendicular directions. Therefore, the term $\delta\bm{E}\cdot\nabla_V\delta f$ is crucial for investigating the time-integrated effect of energy transfer between fields and particles. This term relates directly to the rate of change of the field energy and particles kinetic energy:
\begin{equation}
    \frac{\partial}{\partial t}\left(\frac{\delta B^2}{2\mu_0}\right) \sim-\int \frac{\partial}{\partial t}\left(f_0+\delta f\right)V^{2}d \bm{V} \sim-2\frac{q}{|q|}\frac{\left|q\right|}{m}\int\left(\delta E_{\|} \delta fV_{\|}+\delta E_{\perp 1}\delta f V_{\perp 1}+\delta E_{\perp 2} \delta f V_{\perp 2}\right)d\bm{V}
\end{equation}

Therefore, for $\frac{q}{\left|q\right|}<0$ like in the case of electrons, if $\left<\left(\delta\bm{E}\cdot\bm{V}\right)\delta f\left(\bm{V}\right)\right>>0$, which means that the sense of correlation between $\delta f$ and $\delta\bm{E}$ is the same as the sense of correlation between $\delta\bm{V}$ and $\delta f$, then magnetic field energy increases with time ($\frac{\partial}{\partial t}\left(\left<\frac{\delta B^2}{2\mu_0}\right>\right)>0$).

\subsection{Measurement of the Growth Rate Spectrum}
Following the formula provided by \citet{He2019}, the spectrum of energy conversion rate as a function of frequency reads as
\begin{equation}
    \varepsilon_{\rm JE}=\frac{1}{4}\left(\delta \widetilde{\bm{J}} \cdot \delta \widetilde{\bm{E}}^{*}+\delta \widetilde{\bm{J}}^{*} \cdot \delta \widetilde{\bm{E}}\right)
\end{equation}

where $\widetilde{\bm{J}}$, $\widetilde{\bm{J}}^{*}$, $\widetilde{\bm{E}}$, and $\widetilde{\bm{E}}^{*}$ represent the spectral coefficients and corresponding conjugate counterparts of $\delta J$ and $\delta E$, respectively. Dividing $\varepsilon_{\rm JE}$ by the wave electromagnetic field energy density spectra, we obtain the following growth/damping rate spectrum,
\begin{equation}
    \gamma=-\frac{1}{2}\frac{1}{4}\frac{\left(\delta \widetilde{\bm{J}} \cdot \delta \widetilde{\bm{E}}^{*}+\delta \widetilde{\bm{J}}^{*} \cdot \delta \widetilde{\bm{E}}\right)}{|\delta \widetilde{B}|^{2} / 2 \mu_{0}+\varepsilon_{0}|\delta \widetilde{E}|^{2} / 2}
\end{equation}

\begin{figure}
    \centering
    \includegraphics[width=18cm]{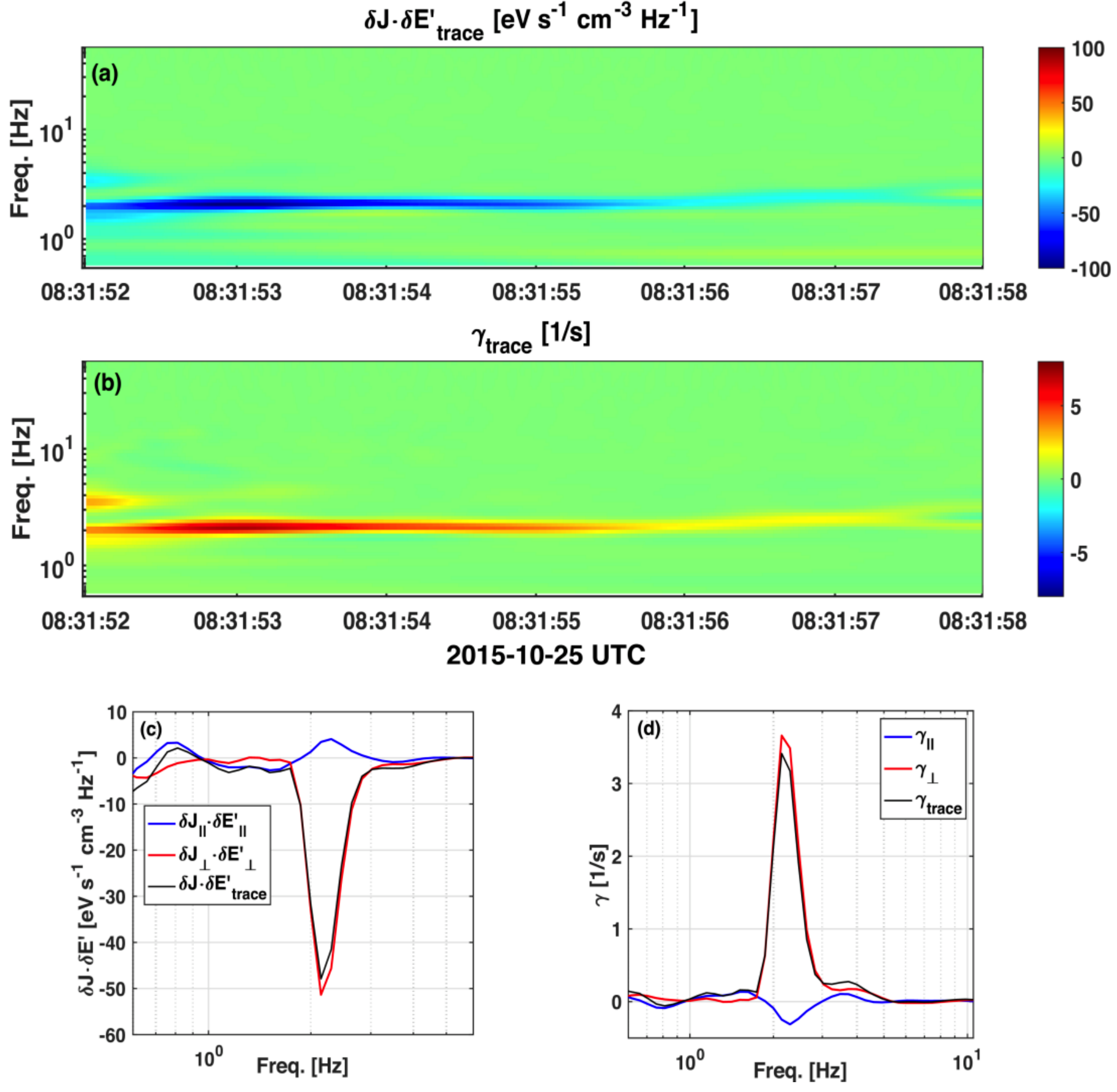}
    \caption{Evidence for ongoing enhancements of wave activity encountered by MMS. (a) Spectrum of $\delta\bm{J}\cdot\delta\bm{E}'$ in a time-period diagram. (b) Spectrum of gamma, normalized growth rate, in the time-period diagram. (c) Frequency profiles of $\delta\bm{J}_{\rm e}\cdot\delta\bm{E}'$ including the trace and the components in the parallel and perpendicular directions. (d) Frequency profiles of gamma (normalized energy conversion rate).}
    \label{fig4}
\end{figure}

The time-frequency spectrum of field-particle energy transfer rate $\varepsilon_{\rm JE}(t, f)$ shows an evident strip near a frequency of 2 Hz (Figure \ref{fig4}a \& \ref{fig4}c). According to the relation $\varepsilon_{\rm JE}(t, f)\sim\gamma\delta B^2$ \citep{He2019}, we obtain the time-averaged growth rate spectra ($\gamma_{\|}(f)$, $\gamma_{\perp}(f)$, and $\gamma_{\rm trace}(f)$) as a function of frequency, manifesting that $\gamma_{\perp}(f)$ and $\gamma_{\rm trace}(f)$ show a distinct bump standing out from the background of the zero growth rate spectrum (Figure \ref{fig4}b \& \ref{fig4}d). We note that the first half of the time interval has larger growth rate and larger amplitude than the second half. Such difference of wave activity between the first and second halves owes to the difference in ion velocity distribution, which has a more pronounced ion-ion drift distribution in the first half.

\section{Theoretical Explanation}
\subsection{Full Set of Eigenmode Solutions Based on Plasma Wave Theory}
We inherit from previous works and develop a new code package called “Plasma Kinetics Unified Eigenmode Solutions” (PKUES) to calculate the eigenmode fluctuations comprehensively under the observed plasma conditions. The set of parameters for the plasma conditions are listed as follows: $\beta_{\rm\parallel c}=6.5$, $T_{\rm\perp c}/T_{\rm\parallel c}=10.2$, $T_{\rm\perp b}/T_{\rm\parallel b}=1.3$, $n_{\rm b}/n_{\rm c}=0.5$, $T_{\rm\parallel b}/T_{\rm\parallel c}=1.1$, $v_{\rm d}/c=-9.5\times10^{-4}$, $w_{\rm\parallel c}/c=2.0\times10^{-4}$, $T_{\rm\parallel e}/T_{\rm\parallel c}=0.7$, $T_{\rm\perp e}/T_{\rm\parallel e}=1.1$, where $\beta_{\rm\parallel c}$ is the proton core-population’s parallel plasma beta value, $T_{\rm\perp c}/T_{\rm\parallel c}$ is the proton core-population’s thermal anisotropy, $T_{\rm\perp b}/T_{\rm\parallel b}$ is the proton beam-population’s thermal anisotropy, $n_{\rm b}/n_{\rm c}$ is the density ratio of proton beam-population to proton core-population, $T_{\rm\parallel b}/T_{\rm\parallel c}$ is the parallel temperature ratio between proton beam-population and proton core-population, $v_{\rm d}/c$ is the core-beam drift speed normalized to the light speed, $w_{\rm\parallel c}/c$ is the proton core-population’s parallel thermal speed as normalized to the light speed, $T_{\rm\parallel e}/T_{\rm\parallel c}$ is the parallel temperature ratio between the electron and proton core population, and $T_{\rm\perp e}/T_{\rm\parallel e}$ is the electron’s thermal anisotropy.

First, we use and modify the dispersion relation solver tool “Plasma Dispersion Relation Kinetics” (PDRK) \citep{Xie2016}, which transforms the dispersion relation to a standard matrix eigenvalue problem of an equivalent linear system, to calculate all the possible solutions of eigenmodes at one time. Our modifications/improvements of PDRK consist of the following three steps: (1) We add the calculation of magnetic helicity and magnetic compressibility to help with the wave mode selection. (2) We distinguish different branches with different colors when displaying the dispersion relations of all wave modes. (3) We realize the function of interactive operation in the part of wave mode selection to select the solutions along the dispersion relation, avoiding manual input of wave model solution.

We then manually choose the most plausible wave mode out of all eigenmode solutions according to the requirement that a selection of critical criteria agree with the observations: type of polarization (e.g., right hand circular polarization), Doppler-shifted wave frequency (i.e., wave frequency in the spacecraft reference frame), transport ratio of fluctuation variables (e.g., large ratio of $\left|\delta E\right|/\left|V_{\rm A}\delta B\right|$), and sign of the growth rate (e.g., positive $\gamma$ for growth process). After determining the most consistent eigenmode with our observations, we then calculate the other fluctuation quantities, e.g., ion and electron fluctuating number densities and bulk velocity vectors, which are not addressed in the PDRK package. Similar to the “New Hampshire Dispersion Solver” (NHDS) \citep{Verscharen2018}, we also calculate the disturbed ion and electron velocity distributions according to Chapter 10 in the book by \citet{Stix1992}. We incorporate the functions of both codes, PDRK and NHDS, and develop a new code package called “Plasma Kinetics Unified Eigenmode Solutions” (PKUES).

\subsection{Similar Field-Particle Correlation Reproduced in Theoretical Model}
\begin{figure}
    \centering
    \includegraphics[width=16cm]{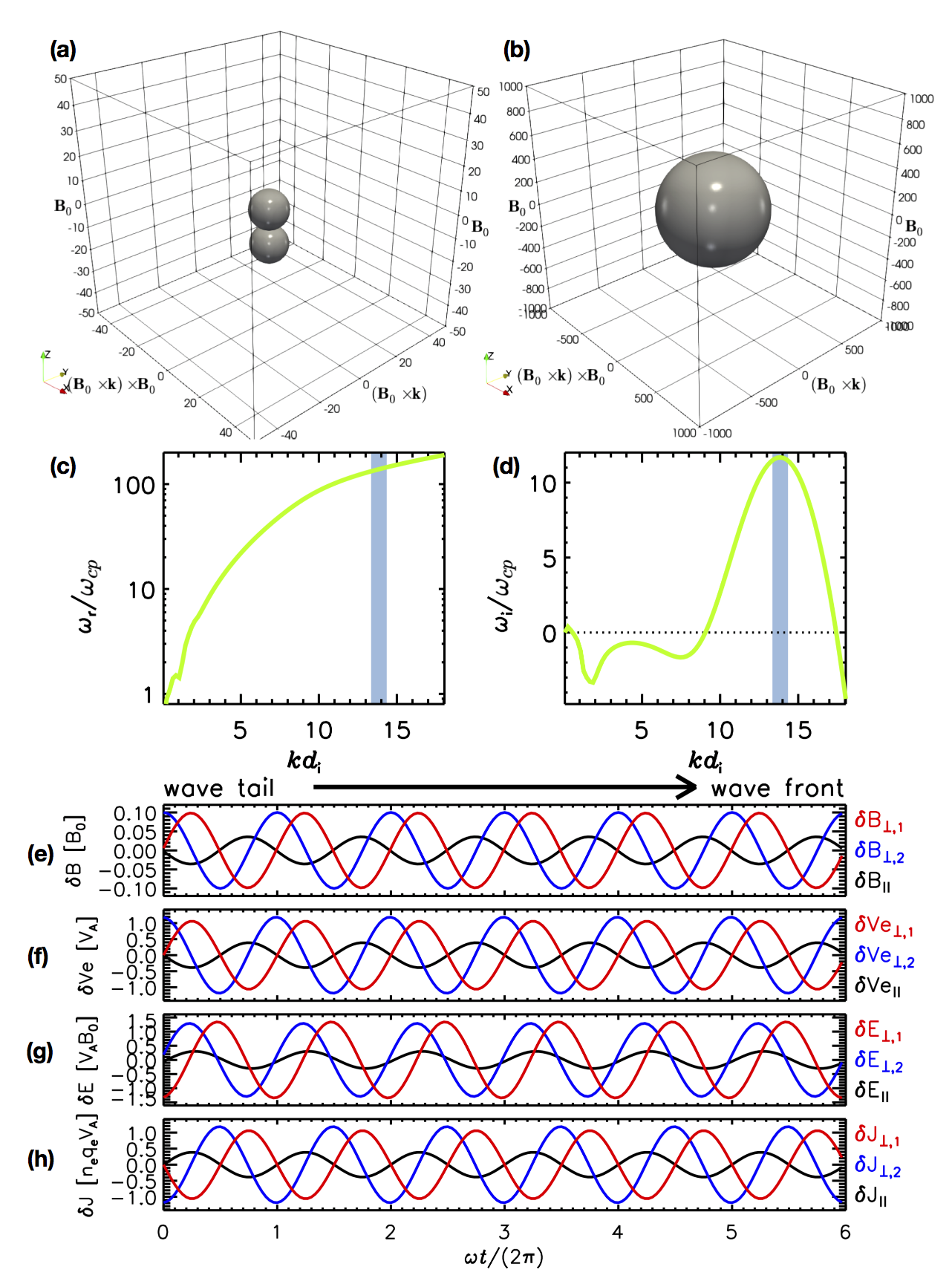}
    \caption{Eigenmode instability driven by non-Maxwellian ion phase space density as predicted from linear theory of plasma kinetics. (a) Ion phase space density with the integrated multi-order moments of the two populations. This theoretical result is similar to the results of our ion observations. (b) Electron phase space density. The integrated multi-order moments are also similar to those of our observations. (c) Dispersion relation and (d) growth rate profile of right-hand polarized whistler waves based on the given ion and electron background velocity distributions. (e-h) Time sequences of the disturbed magnetic field vectors, electron bulk velocity vectors, electric field vectors, and current density vectors.}
    \label{fig5}
\end{figure}

The ion-ion drift instability and the associated excited whistler waves as witnessed by MMS are confirmed with the prediction from plasma wave theory \citep{Gary1991}. Using background plasma parameters consistent with the observations, the ion VDF with an ion-ion core-beam drift and the electron VDF lead to whistler waves in linear theory with the properties as those observed in MMS (Figure \ref{fig5}a \& \ref{fig5}b). Plasma wave theory predicts that the whistler wave is the unstable eigenmode in this case (see Figure \ref{fig5}c \& \ref{fig5}d) for its dispersion and growth rate relations). The wave fluctuations of the magnetic field, electric field, electron bulk velocity, and current density are sampled from the wave tail to the wave head (Figure \ref{fig5}e-\ref{fig5}h) for comparison with the observations in the spacecraft reference frame, where the whistler waves propagate upstream but are convected back by the solar wind flow. 

\begin{figure}
    \centering
    \includegraphics[width=16cm]{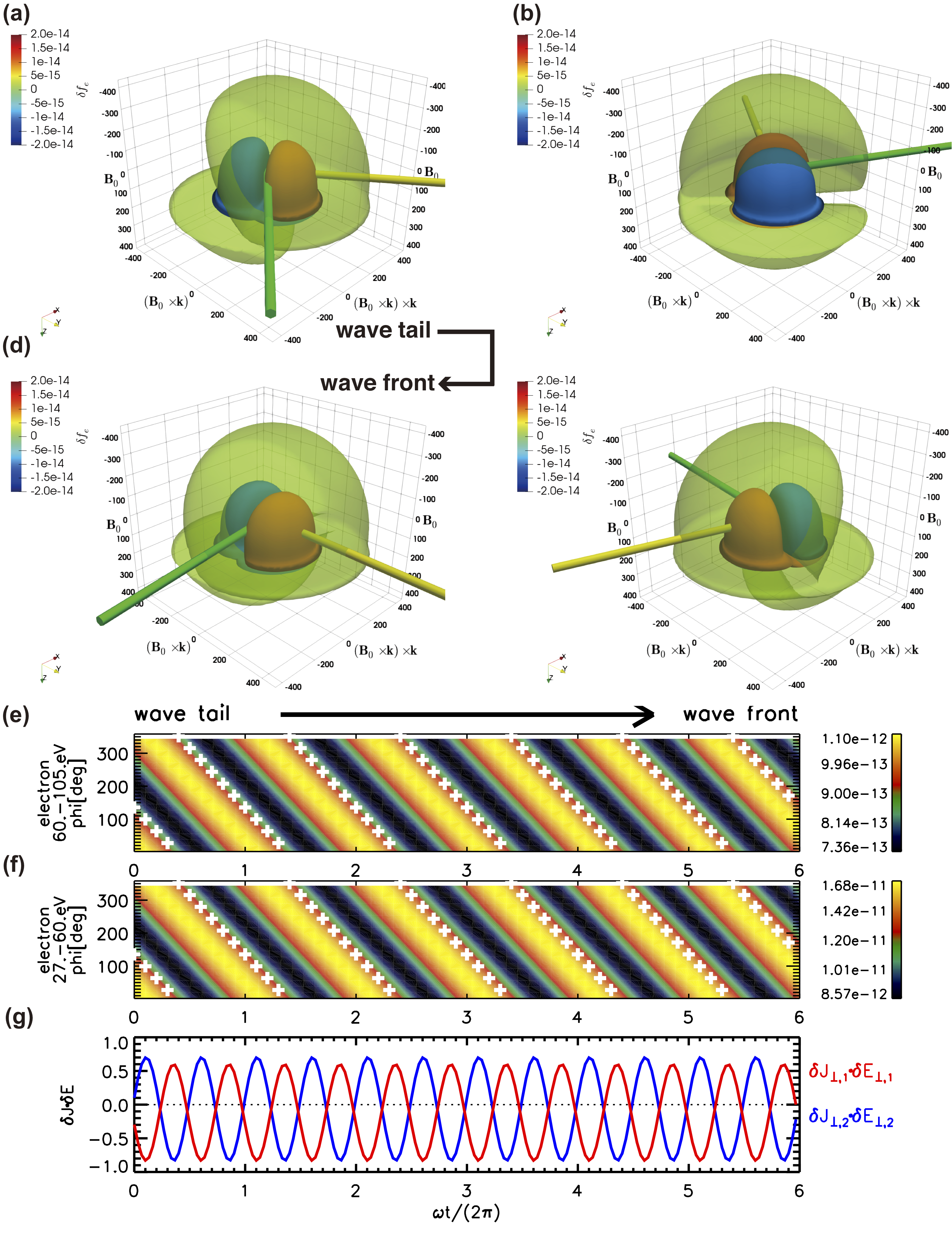}
    \caption{Analysis of field-particle correlations, which are the key underlying concept of secular wave-particle interactions and reponsible for the growth of wave activity. (a-d) Correlations between perturbation of the electron phase space density ($\delta f_{\rm e}$: red isosufrace for $\delta f_{\rm e}>0$, blue isosurface for $\delta f_{\rm e}<0$, green isosurface for $f_{\rm e0}+\delta f_{\rm e}$) and the electromagnetic field vectors (yellow stick for $\delta\bm{E}$, green stick for $\delta\bm{B}$). (e-f) Correlation between the azimuthal angle of the fluctuating electron velocity distribution $\phi_{\rm\delta f_{e}}\left(t\right)$. The angle of a temporally local maximum $\delta f_{\rm e}\left(t,\phi\right)$, and the azimuth angle of the fluctuating electric field $\phi_{\rm\delta E}\left(t\right)$ serve as proxies. (g) Time sequences of the energy conversion rate from particles to electromagnetic fields (red for $\delta\bm{J}_{\rm e,\perp 1}\cdot\delta\bm{E}_{\perp 1}$, blue for $\delta\bm{J}_{\rm e,\perp 2}\cdot\delta\bm{E}_{\perp 2}$).}
    \label{fig6}
\end{figure}

The theoretically predicted phase coherence between the gyrated disturbed electron velocity distribution ($\delta f_{\rm e}$) and the gyrated wave electromagnetic field vectors ($\delta\bm{E}'$ and $\delta\bm{B}$) is of great importance for understanding the relevant field-particle interactions (Figure \ref{fig6}a-\ref{fig6}d, also supplementary Animation 2). Again both the electric and magnetic field vectors ($\delta\bm{E}'$ and $\delta\bm{B}$) point towards positive $\delta f_{\rm e}$, where the angle between $\delta\bm{E}'$ and $\delta\bm{B}$ is smaller than $90^{\circ}$, clearly suggesting the ongoing process of energy transfer from particles to fields and the enhancement of wave electromagnetic field energy. This particle-to-field energy transfer is also confirmed by the azimuthal-angle correlation between the enhanced $f_{\rm e}$ and the wave electric field $\delta\bm{E}'$ (Figure \ref{fig6}e \& \ref{fig6}f), and the time sequence of $\delta\bm{J}\cdot\delta\bm{E}'$ which shows a negative level on average (Figure \ref{fig6}g).

\section{Summary and Discussion}
We take full advantages of the MMS satellite in measuring fields and particles with state-of-art high quality. We witness the growth of whistler waves at a high measurement cadence in space plasmas. As the characteristic perturbations in the unstable whistler waves, we reveal the disturbed electron velocity distribution ($\delta f_{\rm e}$) and find it to be gyrating in phase with the disturbed electromagnetic field vectors, where the field vectors ($\delta \bm{E}'$ and $\delta \bm{B}$) point towards positive $\delta f_{\rm e}$. We conclude that such phase correlations between $\delta f_{\rm e}$ and $\delta \bm{E}'$ are directly evidence of the energy transfer from particles to fields ($\delta\bm{J}\cdot\delta\bm{E}'<0$) that leads to the growth and emission of whistler waves. The wave frequency is estimated to be $\omega_{\rm SC}\sim2.5$ Hz in the spacecraft reference frame, and $\omega_{\rm PL}\sim16$ Hz in the plasma reference frame. The normalized growth rate is comparable in order of magnitude between the direct observation ($\gamma_{\rm obs}\sim3$ Hz) and the theoretical prediction ($\gamma_{\rm theory}\sim1$ Hz). We also obtain the spectrum of growth rate as a function of frequency directly from the in situ measurements. We identify and attribute the free-energy source for the whistler wave’s unstable growth to the proton’s “core-beam-drift” velocity distribution.

\begin{figure}
    \centering
    \includegraphics[width=18cm]{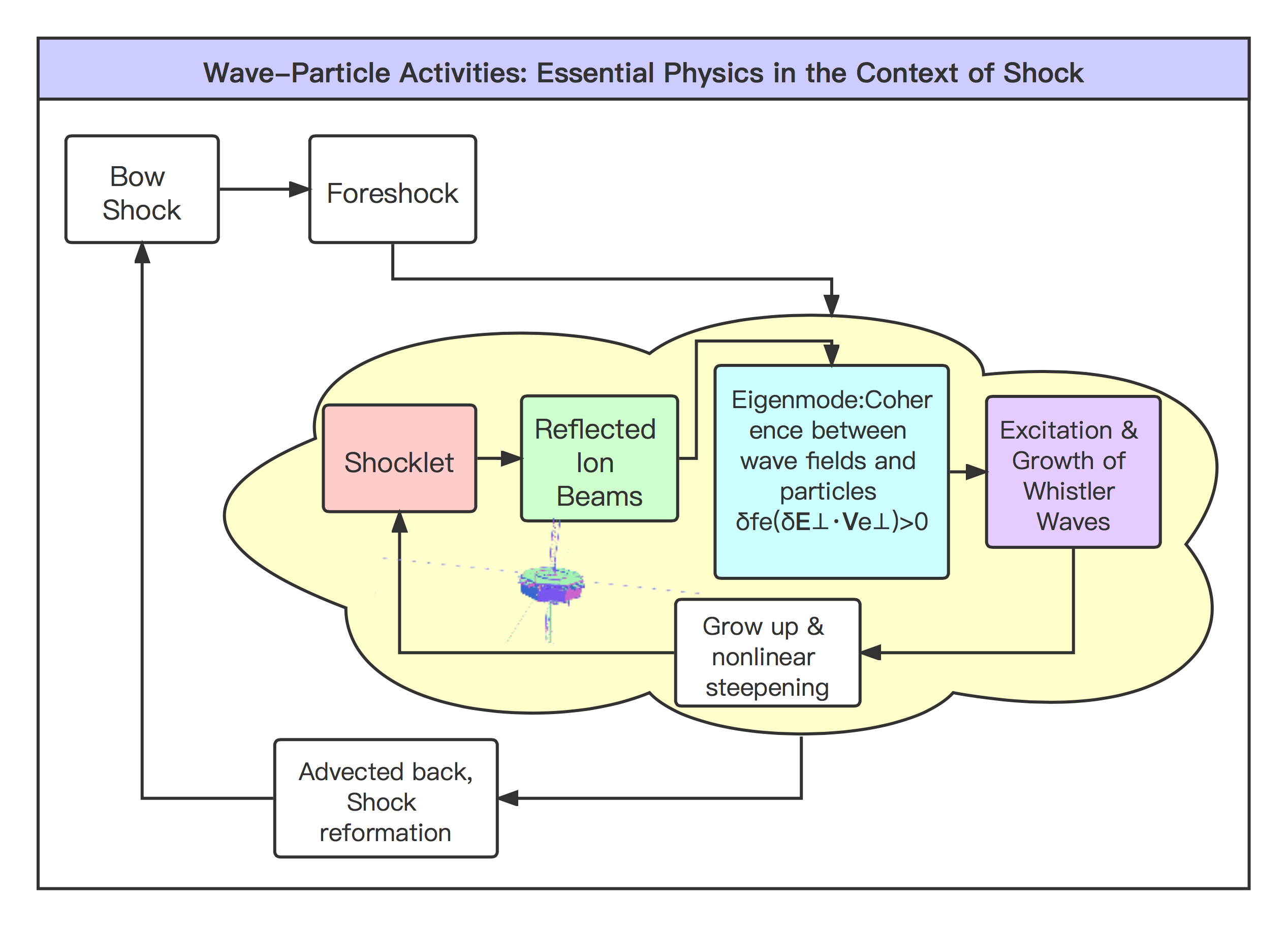}
    \caption{Fundamental physics of field-particle interactions taking place in the foreshock region of planetary bow shock. The drift between solar wind ions and shock-reflected ions in velocity space is intrinsically unstable and determines the dominant eigenmode together that transfers the free energy between fields and particles. Perturbations of the EM-fields and the electron velocity distribution associated with this dominant eigenmode couple to drive the whistler waves, which then evolve into nonlinear large amplitude and may contribute to the reformation of shocklet and shock.}
    \label{fig7}
\end{figure}

The scenario of whistler waves growth revealed in this work (Figure \ref{fig7}) helps to understand the mystery of the origin of whistler waves, which are ubiquitous in various plasma environments. Our observational and theoretical studies are important to define a paradigm on how to explore field-particle couplings and how to quantify the coupling efficiency in collisionless plasmas. In principle, both ions and electrons are linked with and coupled through the electromagnetic fields. However, their dynamic evolution along with the wave fields is distinct from one another in this instability. As the source of free energy, the drifting ion beam population is scattered and decelerated by the Lorentz force as the 2nd order cross product of the disturbed ion beam fluid velocity with the disturbed magnetic field. The scattered and decelerated ion beam would modify the eigenmode’s polarizations, changing the phase relation between $\delta f_{\rm e}$ and $\delta \bm{E}'$ hence saturating the growth of whistler waves. However, the ion beam’s disturbed velocity at the wave frequency cannot be well observed by the MMS plasma instrument, the angular resolution of which makes it difficult to achieve this goal. More advanced space missions dedicated to measuring the solar wind plasma turbulence in the vast heliosphere with an ultra-high angular resolution are required to accomplish the objective of revealing the mysteries of turbulence dissipation and cyclotron wave’s prevalence.


\begin{acknowledgments}
We are grateful to the MMS team for their efforts of operating the measurements, calibrating the raw data, and making the calibrated data available to the public. We also thank the team of 3DView, which is maintained mainly by IRAP/CNES. The work at Peking University is supported by NSFC under contracts 41874200 and 42174194. The work at Sun Yat-sen University is supported by NSFC through grants 41774186. D.V. from UCL is supported by the STFC Ernest Rutherford Fellowship ST/P003826/1 and STFC Consolidated Grant ST/S000240/1. This work is also supported by CNSA under contracts No. D020301 and D020302 and supported by the Key Research Program of the Institute of Geology \& Geophysics, CAS, Grant IGGCAS‐ 201904. About data availability: The MMS dataset analyzed in this work is publicly available at https://lasp.colorado.edu/mms/sdc/public/. About code availability: The code for the tool package of “Plasma Kinetics Unified Eigenmode Solutions” (PKUES) is available at https://github.com/PKU-Heliosphere/PKUES.
\end{acknowledgments}


\begin{thebibliography}{}
\expandafter\ifx\csname natexlab\endcsname\relax\def\natexlab#1{#1}\fi
\providecommand{\url}[1]{\href{#1}{#1}}
\providecommand{\dodoi}[1]{doi:~\href{http://doi.org/#1}{\nolinkurl{#1}}}
\providecommand{\doeprint}[1]{\href{http://ascl.net/#1}{\nolinkurl{http://ascl.net/#1}}}
\providecommand{\doarXiv}[1]{\href{https://arxiv.org/abs/#1}{\nolinkurl{https://arxiv.org/abs/#1}}}

\bibitem[{Bale {et~al.}(2009)Bale, Kasper, Howes, Quataert, Salem, \&
  Sundkvist}]{Bale2009}
Bale, S.~D., Kasper, J.~C., Howes, G.~G., {et~al.} 2009, Phys. Rev. Lett., 103,
  211101, \dodoi{10.1103/PhysRevLett.103.211101}

\bibitem[{{Bruno} \& {Carbone}(2013)}]{Bruno2013}
{Bruno}, R., \& {Carbone}, V. 2013, Living Reviews in Solar Physics, 10, 2,
  \dodoi{10.12942/lrsp-2013-2}

\bibitem[{{Burch} {et~al.}(2016){Burch}, {Moore}, {Torbert}, \&
  {Giles}}]{Burch2016}
{Burch}, J.~L., {Moore}, T.~E., {Torbert}, R.~B., \& {Giles}, B.~L. 2016, \ssr,
  199, 5, \dodoi{10.1007/s11214-015-0164-9}

\bibitem[{{Chen} {et~al.}(2019){Chen}, {Klein}, \& {Howes}}]{Chen2019}
{Chen}, C.~H.~K., {Klein}, K.~G., \& {Howes}, G.~G. 2019, Nature
  Communications, 10, 740, \dodoi{10.1038/s41467-019-08435-3}

\bibitem[{{Eastwood} {et~al.}(2005){Eastwood}, {Lucek}, {Mazelle}, {Meziane},
  {Narita}, {Pickett}, \& {Treumann}}]{Eastwood2005}
{Eastwood}, J.~P., {Lucek}, E.~A., {Mazelle}, C., {et~al.} 2005, \ssr, 118, 41,
  \dodoi{10.1007/s11214-005-3824-3}

\bibitem[{{Fox} {et~al.}(2016){Fox}, {Velli}, {Bale}, {Decker}, {Driesman},
  {Howard}, {Kasper}, {Kinnison}, {Kusterer}, {Lario}, {Lockwood}, {McComas},
  {Raouafi}, \& {Szabo}}]{Fox2016}
{Fox}, N.~J., {Velli}, M.~C., {Bale}, S.~D., {et~al.} 2016, \ssr, 204, 7,
  \dodoi{10.1007/s11214-015-0211-6}

\bibitem[{Gary(1991)}]{Gary1991}
Gary, S.~P. 1991, Space Science Reviews, 56, 373

\bibitem[{{Gershman} {et~al.}(2017){Gershman}, {F-Vi{\~n}as}, {Dorelli},
  {Boardsen}, {Avanov}, {Bellan}, {Schwartz}, {Lavraud}, {Coffey}, {Chandler},
  {Saito}, {Paterson}, {Fuselier}, {Ergun}, {Strangeway}, {Russell}, {Giles},
  {Pollock}, {Torbert}, \& {Burch}}]{Gershman2017}
{Gershman}, D.~J., {F-Vi{\~n}as}, A., {Dorelli}, J.~C., {et~al.} 2017, Nature
  Communications, 8, 14719, \dodoi{10.1038/ncomms14719}

\bibitem[{He {et~al.}(2020)He, Zhu, Verscharen, Duan, Zhao, \& Wang}]{He2020}
He, J., Zhu, X., Verscharen, D., {et~al.} 2020, 898, 43,
  \dodoi{10.3847/1538-4357/ab9174}

\bibitem[{He {et~al.}(2019)He, Duan, Wang, Zhu, Li, Verscharen, Wang, Tu,
  Khotyaintsev, Le, \& Burch}]{He2019}
He, J., Duan, D., Wang, T., {et~al.} 2019, 880, 121,
  \dodoi{10.3847/1538-4357/ab2a79}

\bibitem[{Howes(2017)}]{Howes2017}
Howes, G.~G. 2017, Physics of Plasmas, 24, 055907, \dodoi{10.1063/1.4983993}

\bibitem[{{Kitamura} {et~al.}(2018){Kitamura}, {Kitahara}, {Shoji}, {Miyoshi},
  {Hasegawa}, {Nakamura}, {Katoh}, {Saito}, {Yokota}, {Gershman}, {Vinas},
  {Giles}, {Moore}, {Paterson}, {Pollock}, {Russell}, {Strangeway}, {Fuselier},
  \& {Burch}}]{Kitamura2018}
{Kitamura}, N., {Kitahara}, M., {Shoji}, M., {et~al.} 2018, Science, 361, 1000,
  \dodoi{10.1126/science.aap8730}

\bibitem[{Le {et~al.}(2013)Le, Chi, Blanco-Cano, Boardsen, Slavin, Anderson, \&
  Korth}]{Le2013}
Le, G., Chi, P.~J., Blanco-Cano, X., {et~al.} 2013, Journal of Geophysical
  Research: Space Physics, 118, 2809,
  \dodoi{https://doi.org/10.1002/jgra.50342}

\bibitem[{{Marsch} \& {Verscharen}(2011)}]{Marsch2011}
{Marsch}, E., \& {Verscharen}, D. 2011, Journal of Plasma Physics, 77, 385,
  \dodoi{10.1017/S0022377810000541}

\bibitem[{{Narita}(2018)}]{Narita2018}
{Narita}, Y. 2018, Living Reviewan in Solar Physics, 15, 2,
  \dodoi{10.1007/s41116-017-0010-0}

\bibitem[{{Paschmann} \& {Daly}(1998)}]{Paschmann1998}
{Paschmann}, G., \& {Daly}, P.~W. 1998, ISSI Scientific Reports Series, 1

\bibitem[{{Pollock} {et~al.}(2016){Pollock}, {Moore}, {Jacques}, {Burch},
  {Gliese}, {Saito}, {Omoto}, {Avanov}, {Barrie}, {Coffey}, {Dorelli},
  {Gershman}, {Giles}, {Rosnack}, {Salo}, {Yokota}, {Adrian}, {Aoustin},
  {Auletti}, {Aung}, {Bigio}, {Cao}, {Chandler}, {Chornay}, {Christian},
  {Clark}, {Collinson}, {Corris}, {De Los Santos}, {Devlin}, {Diaz},
  {Dickerson}, {Dickson}, {Diekmann}, {Diggs}, {Duncan}, {Figueroa-Vinas},
  {Firman}, {Freeman}, {Galassi}, {Garcia}, {Goodhart}, {Guererro}, {Hageman},
  {Hanley}, {Hemminger}, {Holland}, {Hutchins}, {James}, {Jones}, {Kreisler},
  {Kujawski}, {Lavu}, {Lobell}, {LeCompte}, {Lukemire}, {MacDonald}, {Mariano},
  {Mukai}, {Narayanan}, {Nguyan}, {Onizuka}, {Paterson}, {Persyn}, {Piepgrass},
  {Cheney}, {Rager}, {Raghuram}, {Ramil}, {Reichenthal}, {Rodriguez},
  {Rouzaud}, {Rucker}, {Saito}, {Samara}, {Sauvaud}, {Schuster}, {Shappirio},
  {Shelton}, {Sher}, {Smith}, {Smith}, {Smith}, {Steinfeld}, {Szymkiewicz},
  {Tanimoto}, {Taylor}, {Tucker}, {Tull}, {Uhl}, {Vloet}, {Walpole}, {Weidner},
  {White}, {Winkert}, {Yeh}, \& {Zeuch}}]{Pollock2016}
{Pollock}, C., {Moore}, T., {Jacques}, A., {et~al.} 2016, \ssr, 199, 331,
  \dodoi{10.1007/s11214-016-0245-4}

\bibitem[{{Santol{\'\i}k} {et~al.}(2003){Santol{\'\i}k}, {Parrot}, \&
  {Lefeuvre}}]{Santolik2003}
{Santol{\'\i}k}, O., {Parrot}, M., \& {Lefeuvre}, F. 2003, Radio Science, 38,
  1010, \dodoi{10.1029/2000RS002523}

\bibitem[{{Shan} {et~al.}(2020){Shan}, {Du}, {Tsurutani}, {Ge}, {Lu},
  {Mazelle}, {Huang}, {Glassmeier}, \& {Henri}}]{Shan2020}
{Shan}, L., {Du}, A., {Tsurutani}, B.~T., {et~al.} 2020, \apjl, 888, L17,
  \dodoi{10.3847/2041-8213/ab5db3}

\bibitem[{{Stansby} {et~al.}(2016){Stansby}, {Horbury}, {Chen}, \&
  {Matteini}}]{Stansby2016}
{Stansby}, D., {Horbury}, T.~S., {Chen}, C.~H.~K., \& {Matteini}, L. 2016,
  \apjl, 829, L16, \dodoi{10.3847/2041-8205/829/1/L16}

\bibitem[{{Stix}(1962)}]{Stix1962}
{Stix}, T.~H. 1962, {The Theory of Plasma Waves}

\bibitem[{{Stix}(1992)}]{Stix1992}
---. 1992, {Waves in plasmas}

\bibitem[{{Torbert} {et~al.}(2016){Torbert}, {Russell}, {Magnes}, {Ergun},
  {Lindqvist}, {Le Contel}, {Vaith}, {Macri}, {Myers}, {Rau}, {Needell},
  {King}, {Granoff}, {Chutter}, {Dors}, {Olsson}, {Khotyaintsev}, {Eriksson},
  {Kletzing}, {Bounds}, {Anderson}, {Baumjohann}, {Steller}, {Bromund}, {Le},
  {Nakamura}, {Strangeway}, {Leinweber}, {Tucker}, {Westfall}, {Fischer},
  {Plaschke}, {Porter}, \& {Lappalainen}}]{Torbert2016}
{Torbert}, R.~B., {Russell}, C.~T., {Magnes}, W., {et~al.} 2016, \ssr, 199,
  105, \dodoi{10.1007/s11214-014-0109-8}

\bibitem[{Tsurutani {et~al.}(1989)Tsurutani, Smith, Brinca, Thorne, \&
  Matsumoto}]{Tsurutani1989}
Tsurutani, B.~T., Smith, E.~J., Brinca, A.~L., Thorne, R.~M., \& Matsumoto, H.
  1989, Planetary and Space Science, 37, 167,
  \dodoi{https://doi.org/10.1016/0032-0633(89)90004-4}

\bibitem[{{Turner} {et~al.}(2020){Turner}, {Liu}, {Wilson}, {Cohen},
  {Gershman}, {Fennell}, {Blake}, {Mauk}, {Omidi}, \& {Burch}}]{Turner2020}
{Turner}, D.~L., {Liu}, T.~Z., {Wilson}, L.~B., {et~al.} 2020, Journal of
  Geophysical Research (Space Physics), 125, e27707,
  \dodoi{10.1029/2019JA027707}

\bibitem[{{Verscharen} \& {Chandran}(2018)}]{Verscharen2018}
{Verscharen}, D., \& {Chandran}, B. D.~G. 2018, Research Notes of the American
  Astronomical Society, 2, 13, \dodoi{10.3847/2515-5172/aabfe3}

\bibitem[{{Verscharen} {et~al.}(2019){Verscharen}, {Klein}, \&
  {Maruca}}]{Verscharen2019}
{Verscharen}, D., {Klein}, K.~G., \& {Maruca}, B.~A. 2019, Living Reviews in
  Solar Physics, 16, 5, \dodoi{10.1007/s41116-019-0021-0}

\bibitem[{{Wilson}(2016)}]{Wilson2016}
{Wilson}, L.~B. 2016, Washington DC American Geophysical Union Geophysical
  Monograph Series, 216, 269, \dodoi{10.1002/9781119055006.ch16}

\bibitem[{{Wilson} {et~al.}(2013){Wilson}, {Koval}, {Szabo}, {Breneman},
  {Cattell}, {Goetz}, {Kellogg}, {Kersten}, {Kasper}, {Maruca}, \&
  {Pulupa}}]{Wilson2013}
{Wilson}, L.~B., {Koval}, A., {Szabo}, A., {et~al.} 2013, Journal of
  Geophysical Research (Space Physics), 118, 5, \dodoi{10.1029/2012JA018167}

\bibitem[{{Xie} \& {Xiao}(2016)}]{Xie2016}
{Xie}, H., \& {Xiao}, Y. 2016, Plasma Science and Technology, 18, 97,
  \dodoi{10.1088/1009-0630/18/2/01}

\bibitem[{Zhao {et~al.}(2019)Zhao, Wang, Shi, Graham, Dunlop, He, Tsurutani, \&
  Wu}]{Zhao2019}
Zhao, J., Wang, T., Shi, C., {et~al.} 2019, 883, 185,
  \dodoi{10.3847/1538-4357/ab3bd1}

\bibitem[{Zong {et~al.}(2020)Zong, Escoubet, Sibeck, Le, \& Zhang}]{Zong2020}
Zong, Q., Escoubet, P., Sibeck, D., Le, G., \& Zhang, H. 2020, Dayside
  Magnetosphere Interactions (Wiley Online Library)

\end{thebibliography}



\end{document}